\documentclass{llncs}
\usepackage{graphicx}
\usepackage{subfigure}
\usepackage{mathtools}
\usepackage{amsmath}
\usepackage{algpseudocode}
\usepackage{algorithm}
\usepackage[export]{adjustbox}
\usepackage{listings}
\usepackage{multirow}
\usepackage{cite}
\usepackage{authblk}
\usepackage{marvosym}
\usepackage{hyperref}
\usepackage{xcolor}
\hypersetup{
    colorlinks=true,
    citecolor=blue,
    linkcolor=blue,
    urlcolor=blue,
}
\usepackage{booktabs}

\begin{document}
\title{General and Practical Tuning Method for Off-the-Shelf Graph-Based Index: SISAP Indexing Challenge Report by Team UTokyo}
\titlerunning{General and Practical Tuning Method for Off-the-Shelf Graph-Based Index}
\author{Yutaro Oguri
\and
Yusuke Matsui
}
\authorrunning{Y. Oguri and Y. Matsui}
\institute{The University of Tokyo, \\ 7-3-1 Hongō, Bunkyo-ku, Tokyo 113-8654, Japan \\
\email{\{oguri,matsui\}@hal.t.u-tokyo.ac.jp}
}
\maketitle
\begin{abstract}
Despite the efficacy of graph-based algorithms for Approximate Nearest Neighbor (ANN) searches, the optimal tuning of such systems remains unclear. This study introduces a method to tune the performance of off-the-shelf graph-based indexes, focusing on the dimension of vectors, database size, and entry points of graph traversal. We utilize a black-box optimization algorithm to perform integrated tuning to meet the required levels of recall and Queries Per Second (QPS). We applied our approach to Task A of the SISAP 2023 Indexing Challenge and got second place in the 10M and 30M tracks. It improves performance substantially compared to brute force methods. This research offers a universally applicable tuning method for graph-based indexes, extending beyond the specific conditions of the competition to broader uses.

\end{abstract}
\section{Introduction}
The proliferation of deep learning has amplified the utility of Nearest Neighbor Search (NNS) in finding the closest vector within a set of embedding vectors for various documents. Particularly for million-scale data, the typical choice is Approximate Nearest Neighbor Search (ANNS). While different ANNS methods exist, graph-based techniques are superior in speed and accuracy, given that the data fits in RAM~\cite{cvpr23_tutorial_neural_search}. Renowned graph-based methods like NSG~\cite{Fu2017FastAN} and HNSW~\cite{Malkov2016EfficientAR} are readily available through optimized libraries like \texttt{Faiss}~\cite{johnson2019billion}.

While off-the-shelf graph indexes provide an efficient baseline, performance tuning becomes crucial to meet specific performance requirements. The evaluation of ANNS performance typically revolves around three metrics: accuracy (Recall@k), runtime (Queries Per Second; QPS), and memory usage. In practical scenarios, such as those presented in the SISAP competition, optimizing one metric often comes with constraints on accuracy, runtime, or memory. A performance tuning method for graph indexes under such constraints is non-trivial and remains an open area of investigation.

Our work proposes a practical approach for performance tuning off-the-shelf state-of-the-art graph-based method (e.g., NSG~\cite{Fu2017FastAN}) according to specified accuracy, runtime, and memory requirements. We focus on three key factors: vector dimensionality reduction, database subsampling, and entry point optimization for graph traversal. We employ black-box optimization for parameter tuning. Our method is flexible and adaptable to various datasets and performance demands. 

We participated in Task A in SISAP Indexing Challenge~\cite{sisap2023challengeweb} and got second place in the final score. In Task A, we use the LAION2B dataset~\cite{schuhmann2022laionb} to perform k-nearest neighbor search ($k=10$). The dataset consists of 16-bit float vectors with 768 dimensions. Under the condition of exceeding a recall of 0.9, the faster the search speed, the higher the score you will receive. We use several subsets (300K, 10M, and 30M size) for evaluation. The submitted code is available at \href{https://github.com/mti-lab/UTokyo-sisap23-challenge-submission}{https://github.com/mti-lab/UTokyo-sisap23-challenge-submission}.

We specifically apply our method to optimize the runtime of NSG index~\cite{Fu2017FastAN} within constraints on accuracy and memory usage. NSG index is a graph-based index approximating MRNG (Monotinic Relative Neighborhood Graph)~\cite{Fu2017FastAN} structure. The time complexity of the search is close to logarithmetic time.

This work makes two key contributions.

\begin{enumerate}
    \item Through exhaustive experiments, we demonstrate that dimensions, database size, and the entry point of the graph traversal serve as valuable parameters for performance tuning.
    \item We introduce a practical and universal method for performing constrained optimization in ANN, considering accuracy, runtime, and memory metrics, utilizing black-box optimization.
\end{enumerate}

\section{Preliminary Study and Findings}
\subsection{Preliminary Study} \label{sec:preliminary}
We first evaluate representative types of indexes with subsets of LAION5B~\cite{schuhmann2022laionb} provided in the competition to choose the baseline. The subset size is 300K, and the query set is 10K public queries provided in the competition.~\cite{sisap2023challengeweb}. The evaluated indexes include the brute force approach, graph-based, PQ-based, and IVF-based index. Evaluation metrics are Recall@k, QPS, and memory usage. Let ground truth k-nearest neighbors be $R$ and approximate nearest neighbors be $\hat{R}$. Recall@k is defined by $\frac{|R \cap \hat{R}|}{k}$. QPS is the average number of processed queries per second. Memory usage represents the index's memory footprint.

All implementations utilize indexes provided by \texttt{Faiss}~\cite{johnson2019billion}, a well-optimized ANNS library with \texttt{C++} and \texttt{Python} bindings. We run preliminary experiments on an Intel(R) Xeon(R) Platinum 8259CL CPU @ 2.50GHz with 512GB RAM.

\begin{figure}[t]
\includegraphics[width=0.60\textwidth]{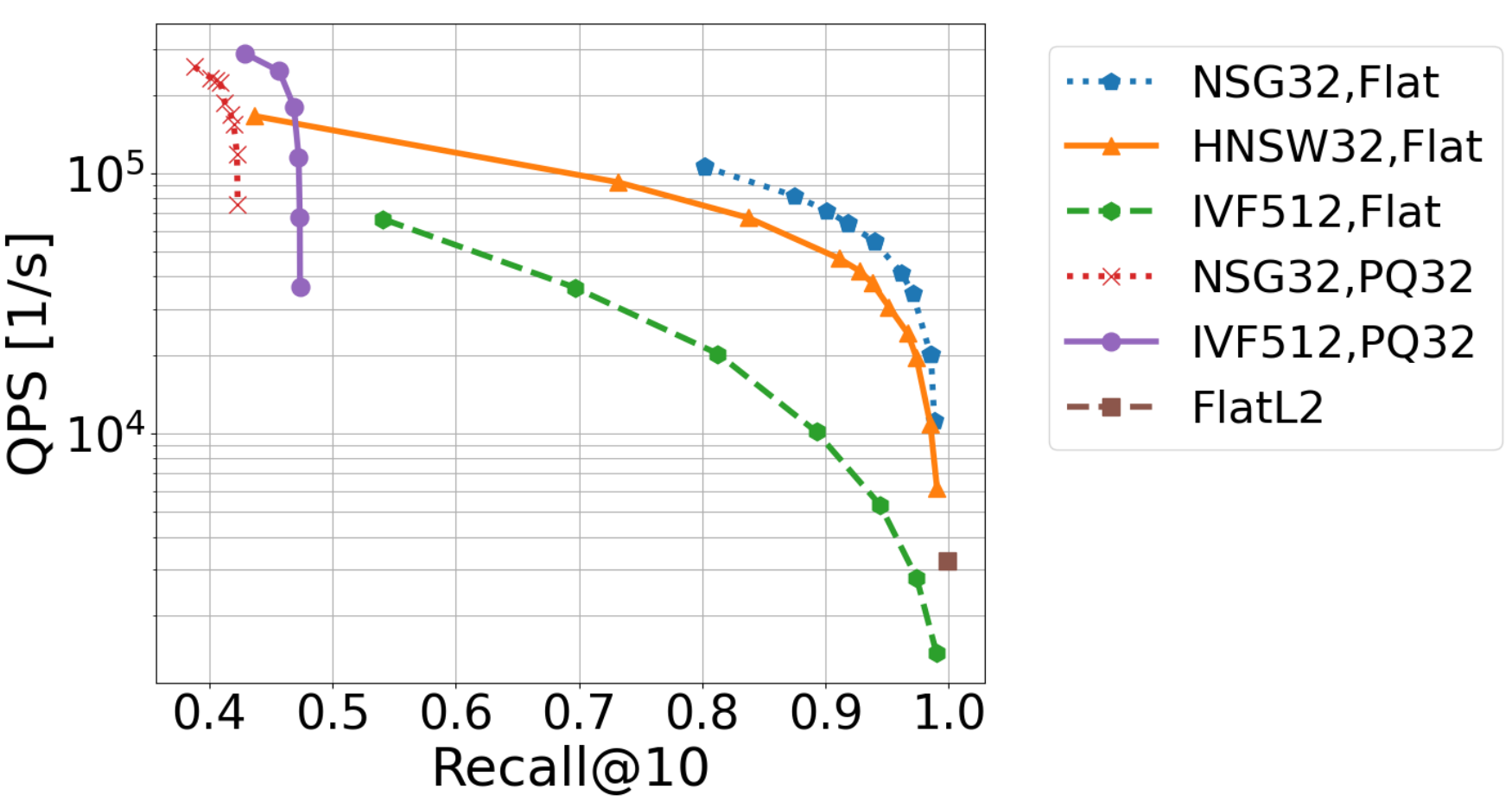}
\centering
\caption{A preliminary experiment comparing various indexes: The FlatL2 means brute-force approach. Other indexes have a common format consisting of two parts separated by a comma. The former means the index name. ``NSG32" means NSG~\cite{Fu2017FastAN} index whose number of links per vertex is 32. (2) ``HNSW32" means HNSW~\cite{Malkov2016EfficientAR} index whose number of links per vertex is 32. (3) ``IVF512" means an inverted file index that divides the dataset into 512 clusters. The latter means the precision of data. ``Flat" means original database vectors, and ``PQ32" means quantized vectors of 32-byte PQ~\cite{PQ2011} code. Note that we did not re-rank the quantized vectors.}
\label{result_preliminary}
\end{figure}

From the outcomes illustrated in Fig.~\ref{result_preliminary}, we found NSG is promising for Task A. We also demonstrate that a graph-based index is the best choice when a memory capacity is sufficient, as often suggested~\cite{cvpr23_tutorial_neural_search}. Among records whose recall is more than 0.9, NSG index runs 22.2 times faster than the brute force method. In addition, despite the memory efficiency and better QPS of the PQ-based approach, we cannot employ it due to its low accuracy. Thus, we select NSG index as a baseline.

In addition, we conducted performance profiling and found that the bottleneck of NSG index is the computation of L2 distances. It occupies a significant fraction (more than 90\%) of the whole computational cost during the search phase. We used \texttt{perf} as a profiling tool.

\subsection{Findings}
Based on these results, we propose three key tuning parameters to improve the search speed: the dimensionality of database vectors, the size of the database, and the entry point for graph traversal. Reducing the dimensionality of vectors and subsampling the database directly reduces the cost of computing L2 distances. In addition, we can change where to start the graph traversal. It is another parameter to be tuned. We aim to optimize these three parameters to improve QPS without compromising Recall@10.

Efficiently selecting these parameters in ANN is not straightforward since Recall@10 and QPS are trade-offs. In addition, the increase in speed due to these parameters is not monotonic due to the various complex factors involved. Moreover, conducting a simple grid search is inefficient. Therefore, this paper proposes a practical framework for tuning an off-the-shelf graph-based index, specifically the NSG index. This framework is a generic method that applies to other types of constraints or different targets.

\section{Method}
\begin{figure}[t]
\includegraphics[width=\textwidth]{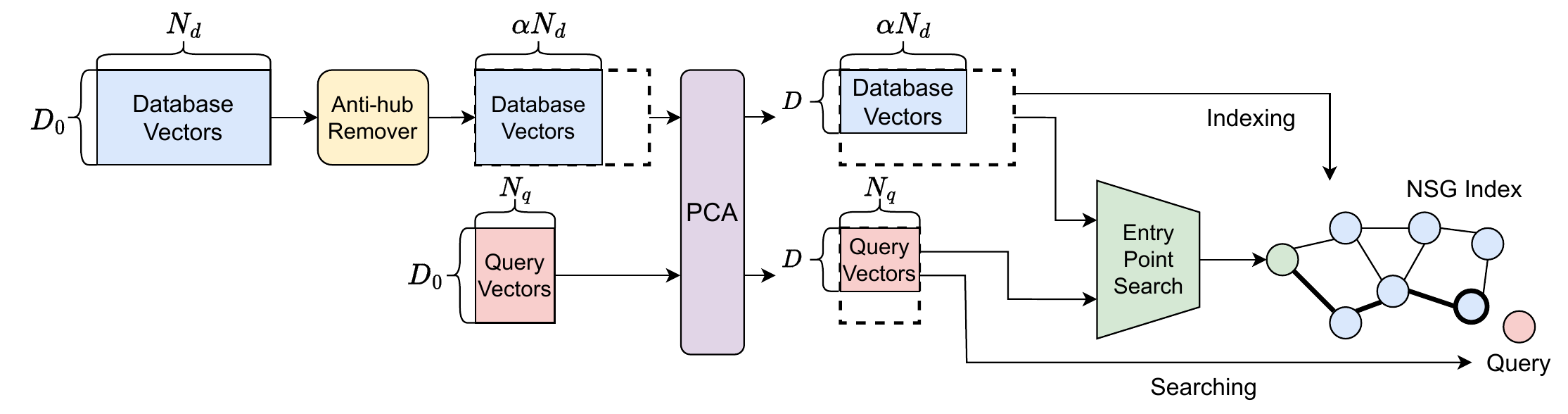}
\caption{The whole pipeline of our method} \label{ann_system}
\end{figure}

Fig.~\ref{ann_system} shows the whole pipeline of our method. It subsamples the database and reduces the dimensionality of vectors. In the search phase, it selects the entry point where the graph traversal begins. We explain the details of each reduction method in Sec.~\ref{sec:component} and how to effectively optimize them in Sec.~\ref{sec:tuning}. Note that we do not modify the graph index itself. Thus, this approach is independent of the implementation of NSG graph index.

\subsection{Components in Pipeline} \label{sec:component}
\subsubsection{Dimensionality Reduction}
We employ Principal Component Analysis (PCA). It is a linear dimensionality reduction algorithm that projects data to a lower dimensional subspace. It reduces the dimension from $D_0$ to $D(\leq D_0)$ (Fig.~\ref{ann_system}). It can directly reduce the computational cost of L2 distance calculation. The reduced dimension $D$ is an indexing parameter to be tuned.

\subsubsection{Database Subsampling}
We employ AntiHub Removal~\cite{antihub2021} to subsample the database effectively. This reduction method is based on hubness in data. It reduce the size of database from $N_d$ to $\alpha N_d (0\leq\alpha\leq 1)$ (Fig.~\ref{ann_system}). It can improve accuracy at a given memory consumption level while maintaining the same QPS. This approach is a compelling tuning candidate because we can apply it with dimensionality reduction methods. The ratio $\alpha$ is also a parameter for indexing.

\subsubsection{Optimizing Entry Point}
If we have multiple entry point candidates, starting with the one closest to the query dramatically speeds up the search~\cite{Arai2021LGTM, iwasaki2018optimization}. We propose a novel and straightforward entry point selection method utilizing $k$-means clustering. It first divides the entire dataset into $k$ clusters and computes a centroid of each cluster (i.e., a centroid is the nearest vector to the mean vector of the cluster). Given a query, we select the closest centroid to the query as an entry point. This approach enables to start traversal from a near point to the query. It prevents excessively long search paths. The number of clusters $k$ is a parameter for building the entry point searcher.

Our approach works well in parallel, even when queries arrive in a batch. \texttt{Faiss} is good at a parallel search at the query level within a batch. However, because our approach requires each query in a batch to have a different optimal entry point, batch processing can become inefficient (Algorithm 1). To address this issue, we propose a gather-style parallel-friendly approach (Algorithm 2). It divides queries into multiple subsets based on optimal entry points and performs batch processing separately for each subset. This approach achieves the same result as Algorithm 1, but with more room for parallel execution (L1 and L7).

\begin{algorithm}
\label{algo:naive}
\caption{An implementation with naive approach}
\begin{lstlisting}[language=Python, numbers=left, xleftmargin=0.58cm]
for query_id, query in enumerate(queries):
    ep = search_entrypoint(query)
    set_entrypoint(index, ep)
    # single query
    results[query_id] = index.search(query, k)
\end{lstlisting}
\end{algorithm}

\begin{algorithm}
\label{algo:improved}
\caption{An implementation for query batch}
\begin{lstlisting}[language=Python, numbers=left, xleftmargin=0.58cm]
epts = search_entrypoints(queries) # runs in batch
for ep in np.unique(epts):
    query_ids = (epts == ep)
    query_batch = queries[query_ids, :]
    set_entrypoint(index, ep)
    # runs in batch
    results[query_ids] = index.search(query_batch, k)
\end{lstlisting}
\end{algorithm}

\subsection{Parameter Tuning with Black-box Optimization} \label{sec:tuning}
We apply a black-box optimization technique to tune parameters $D$, $\alpha$, and $k$ to maximize QPS under memory usage constraints and Recall@$10$, as specified in Task A. As we cannot compute the gradient of QPS with respect to the tunable parameters, we employ black-box optimization. It is an optimization method that does not need derivatives. We use Optuna~\cite{optuna_2019}, a framework for black-box optimization, to implement it. Optuna offers various efficient optimization algorithms. We explore two different strategies under constraints: 1) single-objective optimization with constraint and 2) multi-objective optimization.

\subsubsection{Single-objective Optimization with Constraint}
Single-objective optimization with constraint is formulated as shown in Eqs. 1 and 2. Optuna has a sampler that narrows the parameter search space considering optimization history. TPE (Tree-structured Parzen Estimator) sampler~\cite{TPE2011} supports this type of optimization. It is important to note that it does not guarantee that the obtained solution will always satisfy the constraints; we can only treat them as soft constraints.

\begin{align} \label{eq:single_opt}
    &\text{maximize } \mathrm{QPS} \\
    &\text{subject to } \mathrm{Recall@}k \geq 0.9.
\end{align}

\subsubsection{Multi-objective Optimization}
Multi-objective optimization is formulated as shown in Eq.~\ref{eq:multi_opt}. It can include multiple objective functions, and each of them is desired to be maximized or minimized. TPE sampler~\cite{TPE2011} also supports it. The result is a Pareto frontier, a set of parameter points that achieve the best trade-offs. Since QPS and Recall@k are competing objectives, we can apply multi-objective optimization for Task A. 

\begin{align} \label{eq:multi_opt}
    &\text{maximize } \mathrm{QPS}, \mathrm{Recall@}k.
\end{align}

\section{Experiment}
We evaluate the impact of each of these components on QPS and Recall@k. Then, we conduct an experiment to tune everything integratively with Optuna. We used the same dataset and query set as Sec.~\ref{sec:preliminary}. The tested subset size is 300K, 10M, and 30M. The whole experiments are conducted on the same environment as Sec.~\ref{sec:preliminary}.
\subsection{Ablation Study}

\begin{figure}[t]
    \centering
    \subfigure[]{\includegraphics[width=0.31\textwidth]{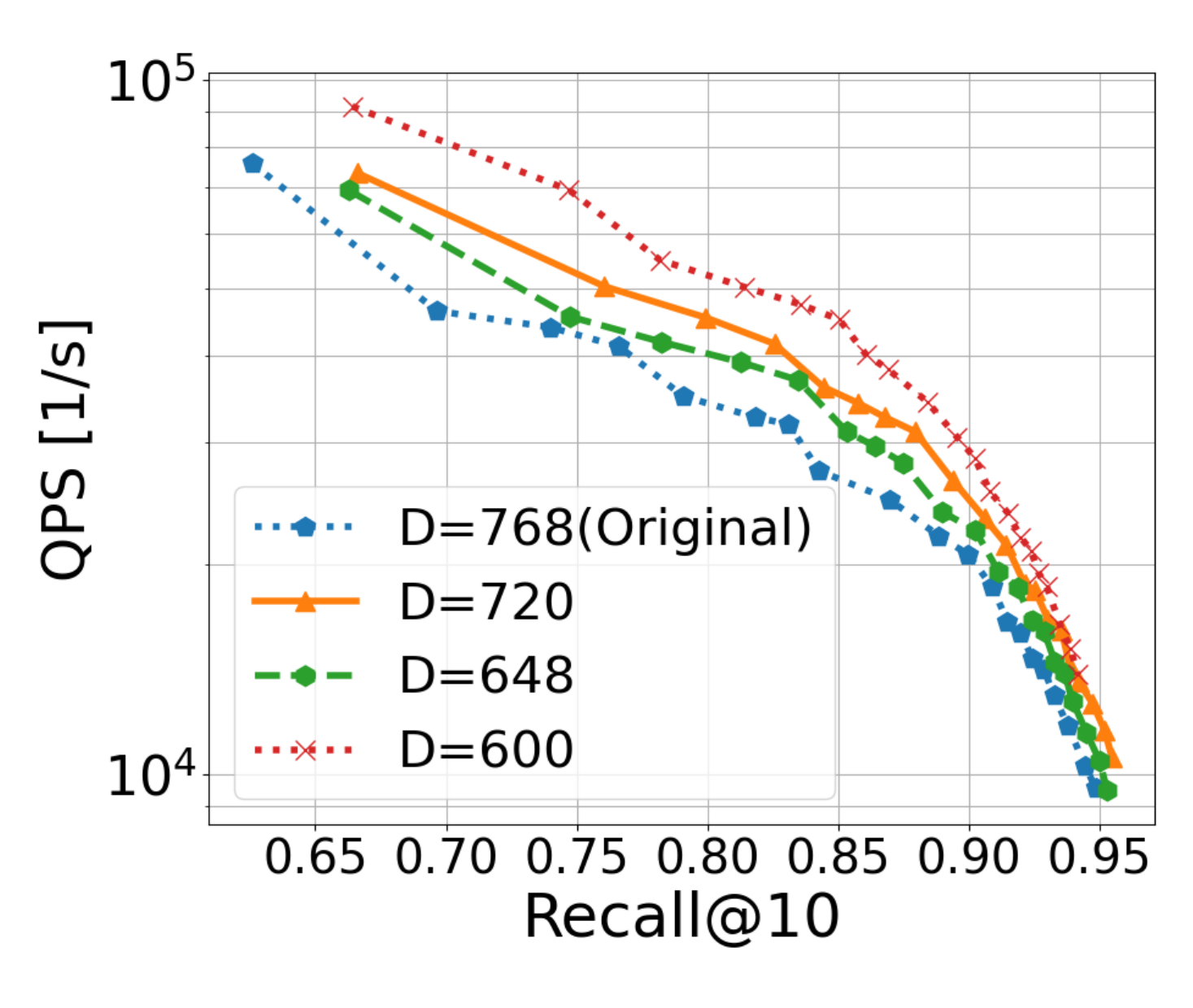}}
    \subfigure[]{\includegraphics[width=0.32\textwidth]{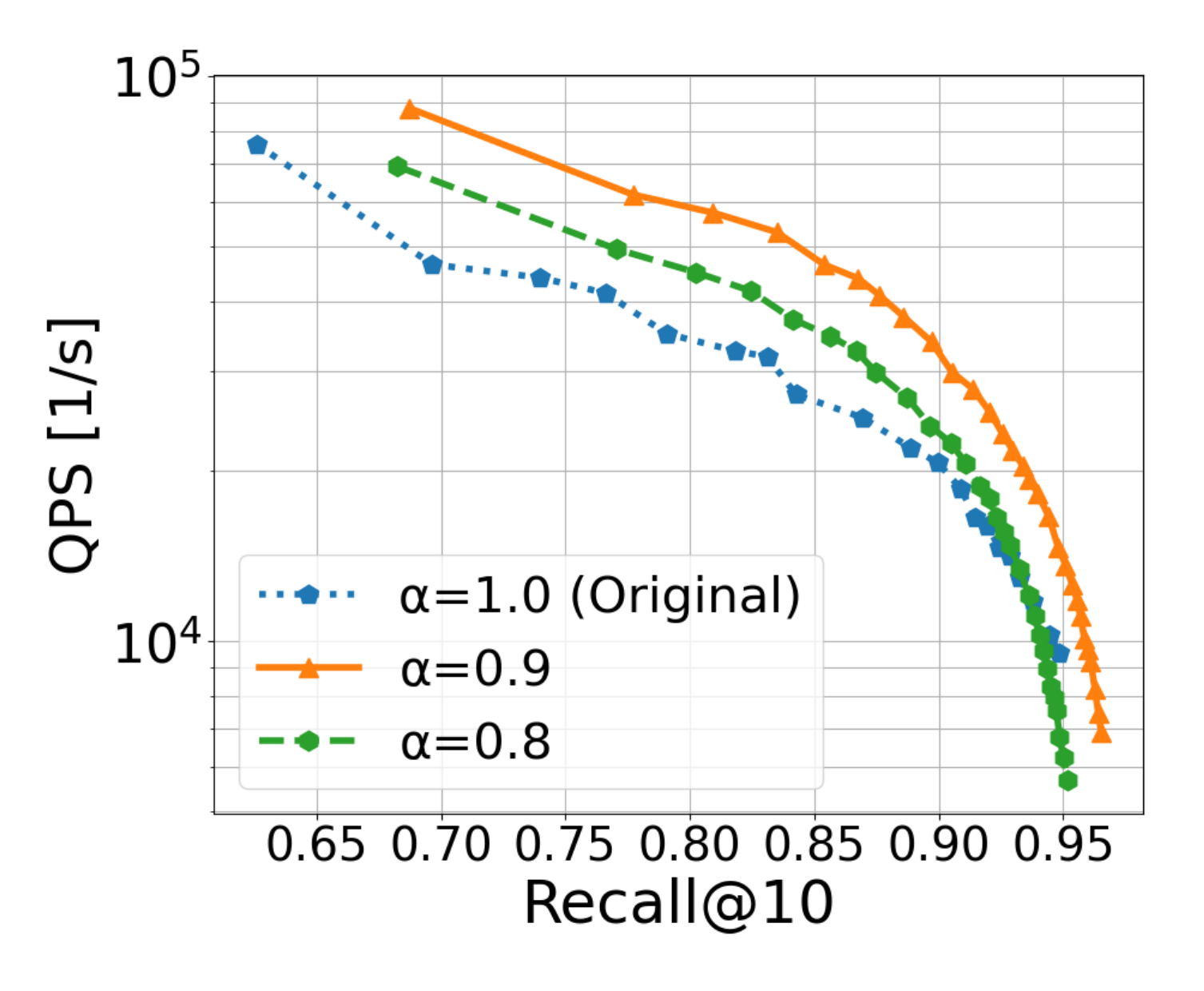}}
    \subfigure[]{\includegraphics[width=0.33\textwidth]{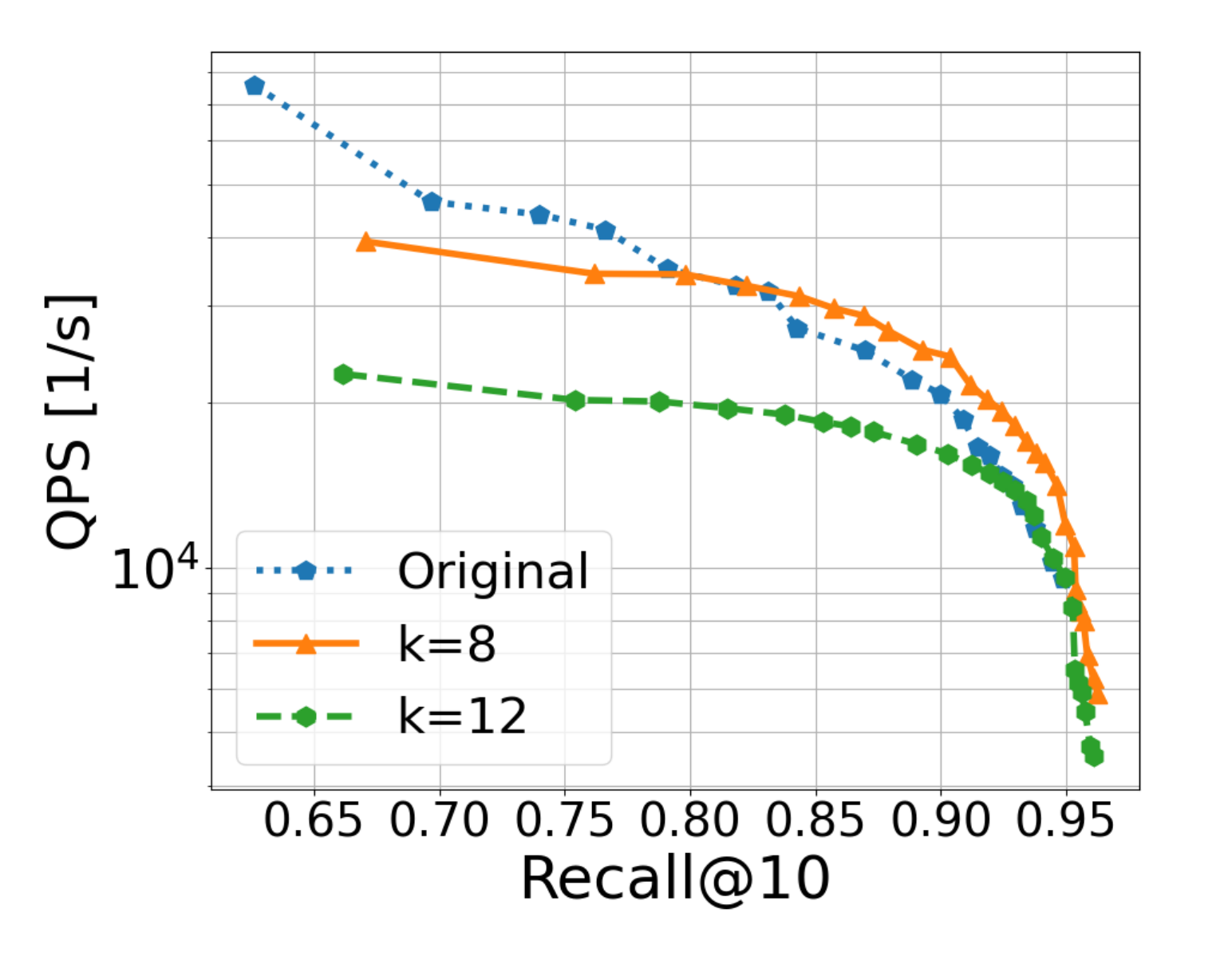}}
    \caption{
        Ablation Study for Each Components (30M subset): 
        (a) PCA + NSG~\cite{Fu2017FastAN}, (b) Antihub Removal~\cite{antihub2021} + PCA and (c) entry point Search with $k$-means + NSG
    }
    \label{fig:ablation}
\end{figure}

\subsubsection{Dimensionality Reduction + NSG (ours) vs Vanilla NSG}
In addition to the vanilla NSG, we apply PCA for dimensionality reduction. We varied the reduced dimension $D$ and measured QPS and Recall@k. The results shown in Fig.~\ref{fig:ablation} (a) demonstrates that applying PCA can increase QPS without compromising accuracy. The best configuration with the condition of $\mathrm{Recall@}k \geq 0.9$ is $D=600$. Its QPS is 1.53 times greater than the best records in the vanilla NSG~\cite{Fu2017FastAN}.

\subsubsection{Subsampling + NSG (ours) vs Vanilla NSG}
We apply Antihub Removal to reduce the size of the database. Fig.~\ref{fig:ablation} (b) shows the performance with various subsampling ratios $\alpha$. It demonstrates that applying subsampling to the database improves efficiency while maintaining accuracy. The best configuration among them is $\alpha=0.9$, which exhibits 1.61 times greater QPS than the vanilla NSG~\cite{Fu2017FastAN}.

\subsubsection{Entry Point Optimization + NSG (ours) vs Vanilla NSG}
We compare performance among various entry point candidates with $k$-means. Fig.~\ref{fig:ablation} (c) demonstrates that optimizing the entry point with $k$-means can potentially increase the QPS in the high accuracy regime. The best configuration shows 1.30 times greater QPS than the vanilla one, while its Recall@10 is 0.9 or greater.

\subsection{Parameter Tuning}
We conducted parameter tuning with black-box optimization. Our ablation study demonstrates that all three aspects can improve performance, and the trends are consistent across different subset sizes for all tuning components. Therefore, we conducted the tuning using a 300K subset for efficiency.

The result demonstrates that multi-objective optimization outperforms single-objective optimization with constraints. When compared over the same tuning time (about 3.5 hours), the best configuration with the former method is 1.85 times faster than that with the latter.

Table.~\ref{tab:result} shows the best results for each subset. We apply tuned parameters for the subset 300K. We choose the best setting among some records for other subsets. It demonstrates that performances for all subsets significantly increased compared to vanilla NSG~\cite{Fu2017FastAN} and brute-force method.

\begin{table}
  \centering
  \caption{The best results for each subset size ($\mathrm{Recall@}k \geq 0.9$)}\label{tab:result}
  \begin{tabular}{@{}lllll@{}}
    \toprule
    & Recall@$10 (\uparrow)$ &\multicolumn{3}{c}{QPS [1/s] $(\uparrow)$} \\ \cmidrule(r){2-2} \cmidrule(){3-5}
     Size & Ours  & Ours & Vanilla NSG~\cite{Fu2017FastAN} & Brute-force\\ \midrule
   300K & $0.9208$ & $\mathbf{1.104} \times \mathbf{10^5}$ $(\times\mathbf{34.16})$ & $7.186 \times 10^4$ $(\times 22.23)$ & $3.232 \times 10^3$ $(\times 1.0)$ \\
   10M & $0.9082$ & $\mathbf{3.822} \times \mathbf{10^4}$ $(\times\mathbf{1078})$ & $2.881 \times 10^4$ $(\times 812.5)$  & $35.46$ $(\times 1.0)$ \\
   30M & $0.9030$ & $\mathbf{3.010} \times \mathbf{10^4}$ $(\times\mathbf{1188})$ & $1.860 \times 10^4$ $(\times 734.6)$ & $25.32$ $(\times 1.0)$ \\
   \bottomrule
  \end{tabular}
\end{table}

\section{Discussion}
\subsection{Applicability to General Settings}
Our framework is practical in other general ANN problems. If there are more complex constraints than the ones in this work, our method may not be suitable, requiring a more complex approach. However, many real-world ANN tunings are oriented towards improving the three axes - Recall, QPS, and Memory - in a straightforward manner. Thus, our approach is applicable in other settings.

In addition, we need to investigate whether the methods proposed here are adequate for graph indexes other than NSG. Since the search for the entry point and the reduction of dimensionality and database are not techniques bound by the specific circumstances of NSG, we can expect their applicability.

\subsection{Comparison to a previous work}
SimilaritySearch.jl~\cite{tellez2022OSSJulia} also introduces an autotuning method for graph-based indices, leveraging a beam search algorithm for parameter tuning~\cite{tellez2022juliav2}. Like our methodology, it models the problem as a black-box optimization to optimize recall and efficiency. SimilaritySearch.jl uses the count of distance computations as its efficiency metric. In contrast, our approach models efficiency using an average QPS measured ten times.
A shared limitation for both methods is the presumption of consistent query distributions during tuning and search. If the assumption is invalid, it might lead to suboptimal outcomes or drastic performance drops.

\subsection{Limitation and Future Work}
Our method cannot satisfy the memory constraint with a 100M subset. It requires further dimensionality and data size reduction, but the problem is that it takes far more time to tune it.

We select conservative parameters to satisfy accuracy for unknown queries in our 10M and 30M submissions. We recognize the need for using more diverse query sets other than public queries for tuning to ensure robust performance.

Lastly, we only used the 300K subset for tuning in these experiments, as the impact of the three parameters we tuned showed consistent trends across all subset sizes. Although it would be ideal to perform tuning on larger subsets, it is exceedingly time-consuming when working with larger subsets. This is because we have to rebuild the index every time the parameters $D$ and $\alpha$ change with each trial. We need to explore practical strategies to reduce the duration.

\section{Conclusion}
In conclusion, this study proposes a successful tuning method for an off-the-shelf graph-based ANN index. By adjusting vector dimension, database size, and graph traversal entry points and utilizing a black-box optimization, we significantly improve Recall@k and QPS performance. We applied our approach to the SISAP Indexing Challenge and significantly outperformed brute force methods. It is also applicable under general conditions.

\section{Acknowledgements}
This work was supported by JST AIP Acceleration Research JPMJCR23U2, Japan.

\bibliographystyle{splncs04}
\bibliography{list}
\end{document}